\documentclass[12pt, notitlepage]{article}
\usepackage{latexsym}
\usepackage[dvips]{graphics,color}
\def\eprint #1 {$\langle$e-print archive: #1$\rangle$}

\font\openface=msbm10 at12pt	% this 12pt might or might not work right with
\def\Integers{\hbox{\openface Z}}

\def\implies{\Rightarrow}
\def\=>{\Rightarrow}
\def\==>{\Longrightarrow}

 \def\dal{\displaystyle{{\hbox to 0pt{$\sqcup$\hss}}\sqcap}}

\def\lto{\mathop
        {\hbox{${\lower3.8pt\hbox{$<$}}\atop{\raise0.2pt\hbox{$\sim$}}$}}}
\def\gto{\mathop
        {\hbox{${\lower3.8pt\hbox{$>$}}\atop{\raise0.2pt\hbox{$\sim$}}$}}}
\def\less{\backslash}		% symbol for set-theoretic difference

\def\to{\rightarrow}		% symbol for `approaches' or `maps to'

\def\tilde{\widetilde}		% define tilde to always be the ``widetilde'' 
\def\ideq{\equiv}		% triple equal sign

\def\interior #1 {  \buildrel\circ\over  #1}     % seems to work

\def\union{\cup}

\def\poscau{{\cal P}}
\def\pp{\varpi}
\def\be{\begin{displaymath}}
\def\ee{\end{displaymath}}
\def\bne{\begin{equation}}
\def\ene{\end{equation}}
\def\bee{\begin{eqnarray*}}
\def\eee{\end{eqnarray*}}
\def\C{\mathcal{C}}

\def\qed{\; \; \Box}
\def\past{\mathrm{past}}

\def\wedge{\,\begin{picture}(2,2) % Lambda
\thicklines
\put(1,2){\circle*{.6}}
\multiput(0,0)(2,0){2}{\circle*{.6}}
\put(0,0){\line(1,2){1}}
\put(2,0){\line(-1,2){1}}
\end{picture}\,\,}
\def\chain3{\,\begin{picture}(0,2) % 3-chain
\thicklines
\multiput(0,0)(0,1){3}{\circle*{.6}}
\put(0,0){\line(0,1){2}}
\end{picture}\,\,}
\def\Lcauset{\,\begin{picture}(2,2) % `L'
\thicklines
\multiput(0,0)(1.5,0){2}{\circle*{.6}}
\put(0,2){\circle*{.6}}
\put(0,0){\line(0,1){2}}
\end{picture}\,\,}
\def\V{\,\begin{picture}(2,2) % V
\thicklines
\put(1,0){\circle*{.6}}
\multiput(0,2)(2,0){2}{\circle*{.6}}
\put(1,0){\line(-1,2){1}}
\put(1,0){\line(1,2){1}}
\end{picture}\,\,}
\def\3antichain{\,\begin{picture}(2,0) % 3-antichain
\thicklines
\multiput(0,.5)(1,0){3}{\circle*{.6}}
\end{picture}\,\,}
\def\wedgeo{\,\begin{picture}(3,2) % Lambda + 1
\thicklines
\put(1,2){\circle*{.6}}
\put(3,0){\circle*{.6}}
\multiput(0,0)(2,0){2}{\circle*{.6}}
\put(0,0){\line(1,2){1}}
\put(2,0){\line(-1,2){1}}
\end{picture}\,\,\,}
\def\No{\,\begin{picture}(3,2) % N + 1
\thicklines
\put(3,0){\circle*{.6}}
\multiput(0,0)(2,0){2}{\circle*{.6}}
\multiput(0,2)(2,0){2}{\circle*{.6}}
\put(0,0){\line(0,1){2}}
\put(0,0){\line(1,1){2}}
\put(2,0){\line(0,1){2}}
\end{picture}\,\,\,}
\def\twoach{\,\begin{picture}(1,1) % ``2-antichain''
\thicklines
\multiput(0,.5)(1,0){2}{\circle*{.6}}
\end{picture}\,\,\,}

\newtheorem{lemma}{Lemma}

\begin{document}
\title{A Classical Sequential Growth Dynamics for Causal Sets\footnote%
{Published in {\it Phys. Rev. D} {\bf 61}, 024002 (2000).
 E-print archive: gr-qc/9904062}}

\author{D. P. Rideout \thanks{rideout@physics.syr.edu} \, and 
	R. D. Sorkin \thanks{sorkin@physics.syr.edu} \\
	Department of Physics, Syracuse University\\ 
	Syracuse, NY, 13244-1130}
\maketitle

%: abstract

\begin{abstract}
\noindent
Starting from certain causality conditions and a discrete form of
general covariance, we derive a very general family of classically
stochastic, sequential growth dynamics for causal sets.  The resulting
theories provide a relatively accessible ``half way house'' to full
quantum gravity that possibly contains the latter's classical limit
(general relativity).  Because they can be expressed in terms of state
models for an assembly of Ising spins living on the relations of the
causal set, these theories also illustrate how non-gravitational matter
can arise dynamically from the causal set without having to be built in
at the fundamental level.  Additionally, our results bring into focus
some interpretive issues of importance for causal set dynamics, and for
quantum gravity more generally.
\end{abstract}

\section{Introduction}
The causal set hypothesis asserts that spacetime, ultimately, is
discrete and that its underlying structure is that of a locally
finite, partially ordered set (a causal set).  The approach to quantum
gravity based on this hypothesis has experienced considerable progress
in its kinematic aspects.  For example, one possesses natural
extensions of the concepts of proper time and spacetime dimensionality
to causal sets, and these take us a significant way toward an answer
to the question, ``When does a causal set resemble a Lorentzian
manifold?''.  The dynamics of causal sets (the ``equations of
motion''), however, has not been very developed to date.  One of the
primary difficulties in formulating a dynamics for causal sets is the
sparseness of the fundamental mathematical structure.  When all one
has to work with is a discrete set and a partial order, even the
notion of what we should mean by a dynamics is not obvious.

Traditionally, one prescribes a dynamical law by specifying a
Hamiltonian to be the generator of the time evolution.  This practice
presupposes the existence of a continuous time variable, which we do
not have in the case of causal sets.  Thus, one must conceive of
dynamics in a more general sense.  In this paper, evolution will be
envisaged as a process of stochastic growth to be described in terms
of the probabilities (in the classical case, or more generally the
quantum measures in the quantum case \cite{qmeasure}) of forming
designated causal sets.  That is, the dynamical law will be a rule
which assigns probabilities to suitable classes of causal sets (a
causal set being the ``history'' of the theory in the sense of
``sum-over-histories'').  One can then use this rule -- technically a
probability measure -- to ask physically meaningful questions of the
theory.  For example one could ask ``What is the probability that the
universe possesses the diamond poset as a partial stem?''. (The term
`stem' is defined below.)

Why are we interested in a classical dynamics for causal sets, when our
ultimate aim is a quantum theory of gravity?  One obvious reason is that
the classical case, being much simpler, can help us to get used to a
relatively unfamiliar type of dynamical formulation, bringing out the
pertinent physical issues and guiding us toward physically suitable
conditions to place on the theory.  Is there, for example, an
appropriate form of causality that we can impose?  Should we attempt to
express the theory directly in terms of gauge invariant (labeling
independent) quantities, or should we follow precedent by enforcing
gauge invariance only at the end?  Some of these issues are well
illustrated with the theories we construct herein.

One of the best reasons to be interested in a classical dynamics for
causal sets is that quantum gravity must possess general relativity as a
classical limit.  Thus general relativity should be described as some
type of effective classical dynamics for causal sets, and one may hope
that the relevant dynamical law will be among the family delineated
herein.  (One can't be certain this will occur, because general
relativity, as a continuum theory, seems most likely to arise as an
effective theory for coarse-grained causal sets, rather than directly as
a limit of the microscopic discrete theory, and there is no guarantee
that this effective theory will have the same form as the underlying
exact one.)

A question commonly asked of the causal set program is ``How could
nongravitational matter arise from only a partial order?''.  One
obvious answer is that matter can emerge as a higher level construct
via the Kaluza-Klein mechanism \cite{JFKK}, but this possibility has
nothing to do with causal sets as such.  The theory developed herein
suggests a different mechanism.  It is possible to rewrite the theory
in such a way that the dynamics appears to arise from a kind of
``effective action'' for a field of Ising spins living on the
relations of the causal set.  A form of ``Ising matter'' is thus
implicit in what would seem at first sight to be a purely
``source-free'' theory.

In subsequent sections of this paper we: describe our notation and
terminology (some new language is required for the detailed derivation
of our causal set dynamics); introduce and briefly discuss the
transitive percolation model; present the physical requirements of
Bell causality and discrete general covariance that we will impose;
derive the (generically)
most general theory satisfying these requirements
(including solving the inequalities which express
that all probabilities must fall between 0 and 1); single out a few
simple choices of the free parameters and exhibit some properties of
the resulting ``cosmologies''; exhibit a pair of state-models for the
dynamics that illustrate how not only geometry, but other matter can
arise implicitly from order.

\subsection{Notation/Terminology/Sequential growth}
First we establish our terminology and notation.  For a fuller
introduction to causal sets, 
see \cite{causets0,causets1,causets2,causets4,bom87}.
(For recent examples of other discrete models incorporating a causal
ordering see \cite{fot,amb1,amb2}.)

A {\it causal set} (or ``causet'') is a locally finite, partially
ordered set (or ``poset'').  We represent the order-relation by
`$\prec$' and use the {\it irreflexive convention} that an element
does not precede itself.

Let $C$ be a poset.  The {\it past} of an element $x\in{C}$ is the
subset $\past(x)=\{y\in{C}\,|\,y\prec{x}\}$.  The past of a subset of
$C$ is the union of the pasts of its elements.  An element of $C$ is
{\it maximal} iff it is to the past of no other element.  A {\it
chain} is a linearly ordered subset of $C$ (a subset, every two
elements of which are related by $\prec$); an {\it antichain} is a
totally unordered subset (a subset, no two elements of which are
related by $\prec$).  A \emph{partial stem} of $C$ is a finite subset
which contains its own past.  (A full stem is a partial stem such that
every element of its complement lies to the future of one of its
maximal elements.)  An \emph{automorphism} of $C$ is a one-to-one map
of $C$ onto itself that preserves $\prec$.

A \emph{link} of a poset is an irreducible relation, that is, one not
implied by other relations via transitivity.\footnote%
{Links are often called ``covering relations'' in the mathematical
 literature.}
A {\it path} in a poset is an increasing sequence of elements, each
related to the next by a link.

A poset can be represented graphically by a Hasse diagram, which is
constructed as follows.  Draw a dot to represent each element of the
poset.  Draw a line connecting any two elements $x\prec{}y$ related by a
link, such that the preceding element $x$ is drawn below the following
element $y$.

The dynamics which we will derive can be regarded as a process of
``cosmological accretion'' or ``growth''.  At each step of this process
an element of the causal set comes into being as the ``offspring'' of a
definite set of the existing elements -- the elements that form its
past.  The phenomenological passage of time is taken to be a
manifestation of this continuing growth of the causet.  Thus, we do not
think of the process as happening ``in time'' but rather as
``constituting time'', which means in a practical sense that there
is no meaningful order of birth of the elements other than that
implied by the relation $\prec$.

In order to define the dynamics, however, we will treat the births as if
they happened in a definite order with respect to some fictitious
``external time''.  In this way, we introduce an element of ``gauge''
into the description of the growth process which we will have to
compensate by imposing appropriate conditions of ``gauge invariance''.
This fictitious order of birth can be represented as a 
{\it natural labeling} of the elements, that is, a labeling by integers
$0,1,2\cdots{}$ which are compatible with the causal order $\prec$ in
the sense that $x\prec{y}\implies{label}(x)<label(y)$.\footnote%
{A natural labeling of an order $P$ is equivalent to what is called a
``linear extension of $P$'' in the mathematical literature.}
The relevant notion of gauge invariance (which we will call ``discrete
general covariance'') is then captured by the statement that the
labels carry no physical meaning.  We discuss this more extensively
later on.\footnote%
{The continuum analog of a natural labeling might be a coordinate system
 in which $x^0$ is everywhere timelike (and this in turn is almost the
 same thing as a foliation by spacelike slices).  One could also
 consider arbitrary labelings, which would be analogous to arbitrary
 coordinate systems.  In that case, there would be a well-defined gauge
 {\it group} --- the group of permutations of the causet elements ---
 and labeling invariance would signify invariance under this group, in
 closer analogy with diffeomorphism invariance and ordinary gauge
 invariance.  However, we have not found a useful way to do this, and in
 this paper only natural labelings will ever be considered.}

\begin{figure}[htbp]
\center
\scalebox{.61}{\includegraphics{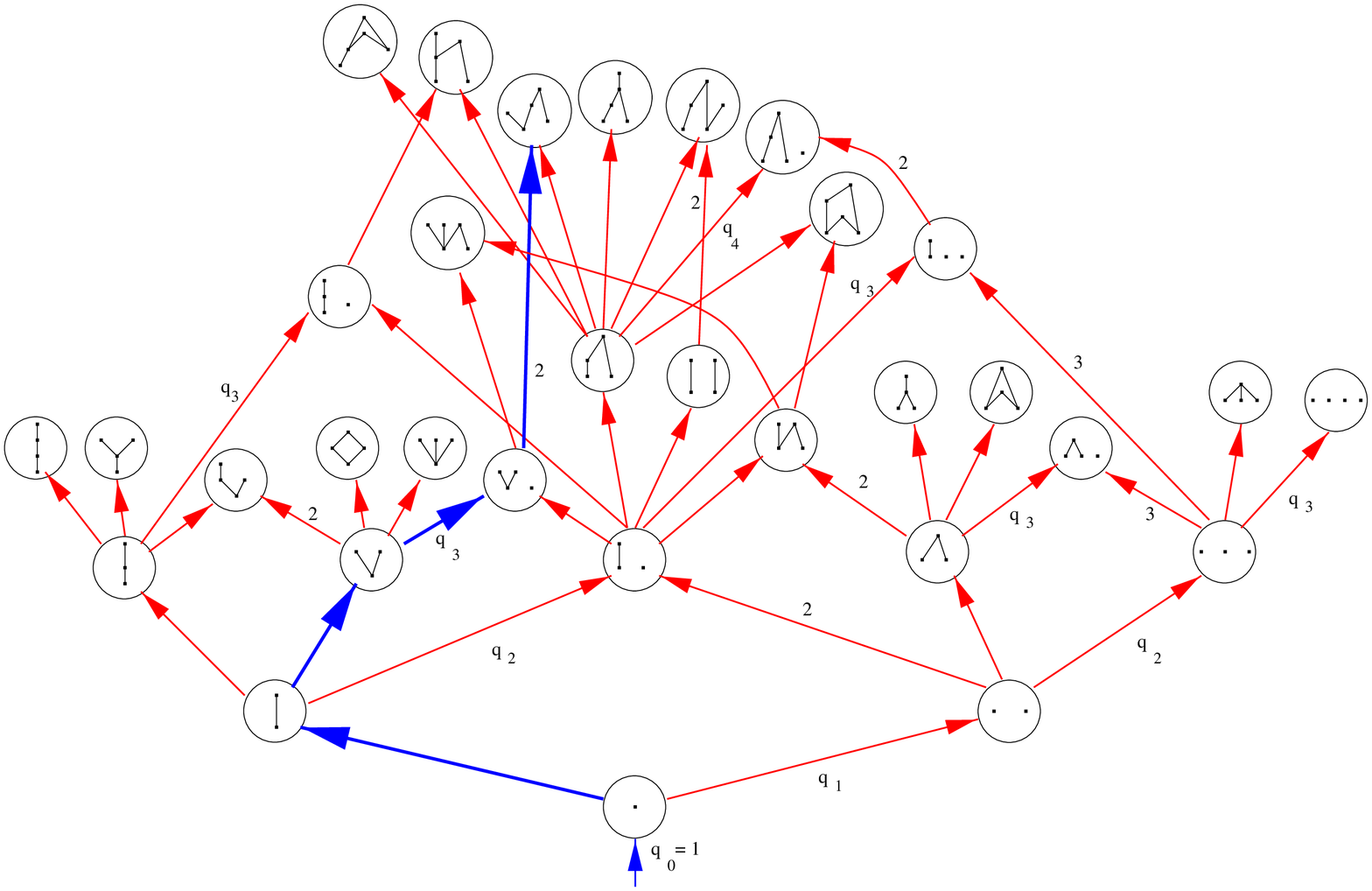}} 
\caption{The poset of finite causets}
\label{poscau}
\end{figure}

It is helpful to visualize the growth of the causal set in terms of
paths in a poset $\poscau$ of finite causal sets.  (Thus viewed, the
growth process will be a sort of Markov process taking place in
$\poscau$.)  
Each finite causet 
(or rather each isomorphism equivalence class of finite causets) 
is one element of this poset.  
If a causet can be formed by accreting a single element to a second causet, 
then the former (the ``child'') follows the latter (the  ``parent'')
in $\poscau$ and the relation between them is a link.  
Drawing $\poscau$ as a Hasse diagram of Hasse diagrams, 
we get figure \ref{poscau}.  
(Of course this is only a portion of the infinite diagram; 
it includes all the causal sets of fewer than five elements 
and 8 of the 63 five element causets.
The ``decorations'' on some of the transitions in figure \ref{poscau} are
for later use.)
Any natural labeling of a causet $C\in\poscau$ determines uniquely a
path in $\poscau$ beginning at the empty causet and ending at $C$.
Conversely, any choice of upward path through this diagram determines 
a naturally labeled causet
(or rather a set of them, since inequivalent labelings can sometimes
give rise to the same path in $\poscau$.)\footnote%
{We could restore uniqueness by ``resolving'' each link $C_1\prec{C_2}$
 of $\poscau$ into the set of distinct embeddings $i:C_1\to{C_2}$ that
 it represents.  Here, two embeddings count as distinct iff no
 automorphism of the child relates them (cf. the discussion of the
 Markov sum rule below).}
We want the physics to be independent of labeling, so different paths in
$\poscau$ leading to the same causet should be regarded as representing
the same (partial) universe, the distinction between them being ``pure
gauge''.

The causal sets which can be formed by adjoining a single 
maximal element to a
given causet will be called collectively a \emph{family}.  The causet
from which they come is their \emph{parent}, and they are
\emph{siblings} of each other.  Each one is a \emph{child} of the
parent.  The child formed by adjoining an element which is to the future
of every element of the parent will be called the \emph{timid child}.
The child formed by adjoining an element 
which is spacelike to every other element 
will be called the \emph{gregarious child}.  
%% A child which is not the timid child will be called a \emph{bold} child.

Each parent-child relationship in $\poscau$ describes a `transition'
$C\to{}C'$, from one causal set to another induced by the birth of a
new element.  The past of the new element (a subset of $C$) will be
referred to as the \emph{precursor set} of the transition (or
sometimes just the ``precursor of the transition'').  Normally, this
precursor set is uniquely determined up to automorphism of the parent
by the (isomorphism equivalence class of the) child, but (rather
remarkably) this is not always the case.  The symbol $\C_n$ will
denote the set of causets with $n$ elements, and the set of all
transitions from $\C_n$ to $\C_{n+1}$ will be called \emph{stage $n$}.

As just remarked, each parent-child transition corresponds to a
choice of partial stem in the parent (the precursor of the transition).
Since there is a one-to-one correspondence between partial stems and
antichains, a choice of child also corresponds to a choice of (possibly
empty) antichain in the parent, the antichain in question being the set
of maximal elements of the past of the new element.  Note also that the
new element will be \emph{linked} to each element of this antichain.

\subsection{Some examples}
\begin{figure}[htbp]
\center
\scalebox{.9}{\includegraphics{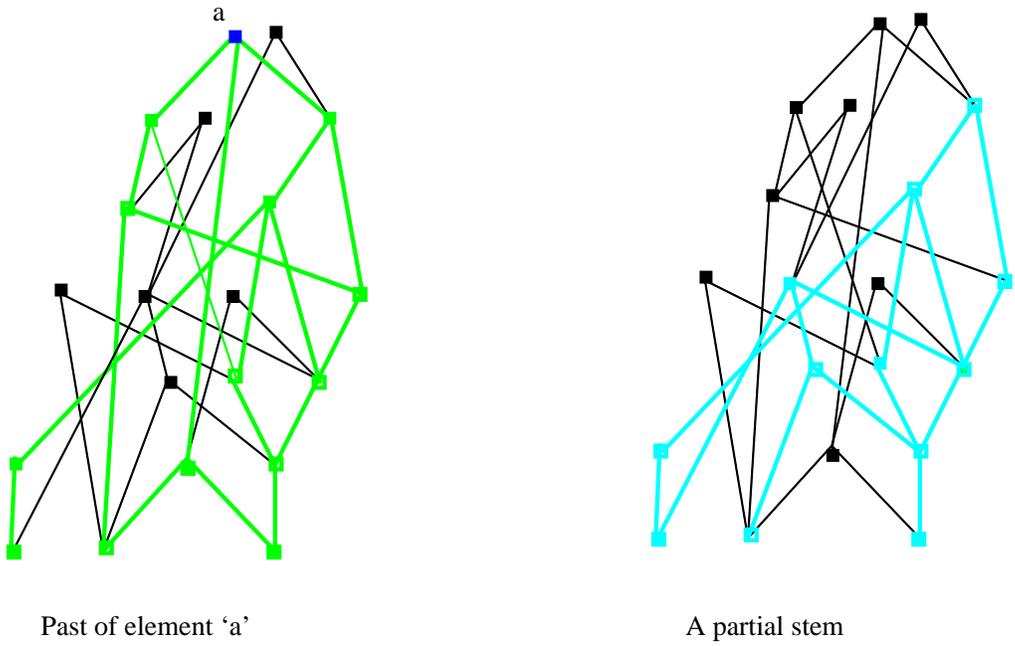}}
\caption{An example of a (`typical'?) 20 element causal set}
\label{20elts}
\end{figure}

To help clarify the terminology introduced in the previous section, we
give some examples.  
The 20 element causet of figure \ref{20elts} 
was generated by the stochastic dynamics described herein, 
with the choice of parameters given by 
equation (\ref{lifelike}) below.
In the copy of this causet on the left, the past of element $a$ is
highlighted.  Notice that since we use the irreflexive convention for
the order, $a$ is not included in its own past.  In the the copy on the
right, a partial stem of the causet is highlighted.

\begin{figure}[htbp]
\center
\scalebox{.64}{\includegraphics{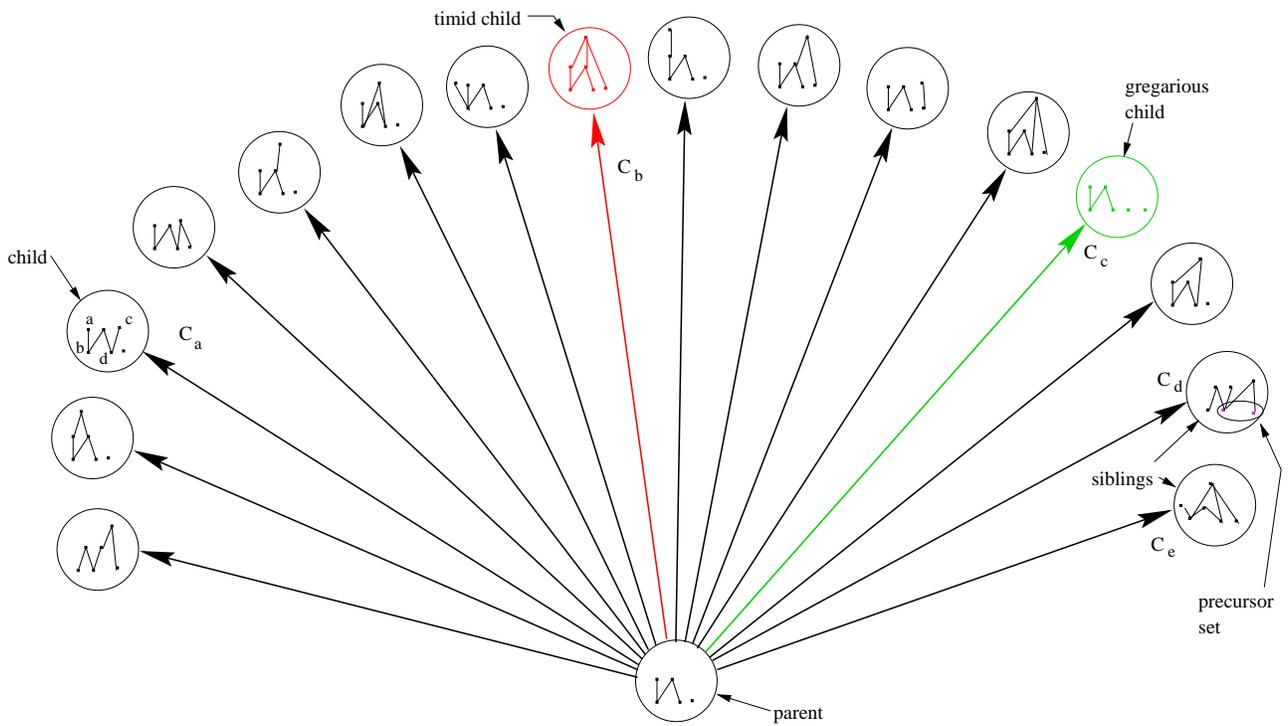}}
\caption{A family}
\label{family}
\end{figure}

Figure \ref{family} shows \No and its children.
The timid child is $C_b$ and the gregarious child is $C_c$.  The
precursor set leading to the transition to $C_d$ is shown in the
ellipse.  An example of an automorphism of $C_a$ is the map
$a\leftrightarrow c, b\leftrightarrow d$ (the other elements remaining
unchanged).

\section{Transitive Percolation}
\label{plain_perc}

In a sum over histories formulation of causal set theory, 
one might expect sums like
\bne
 \sum_\C A(\C,\{q\})
 \label{soh}
\ene 
to be involved, where $A$ is a complex amplitude for the causal set
$\C$, possibly depending on a set of parameters $\{q\}$.  Kleitman and
Rothschild have shown that the number of posets of cardinality $n$ grows
faster than exponentially in $n$ and that asymptotically, almost every
poset has a certain, almost trivial, ``generic'' form.  
(See \cite{brightwell}.)  
Such a ``generic poset'' consists of three ``tiers'', 
with $n/2$ elements in the middle tier 
and $n/4$ elements in the top and bottom tiers.  
For this reason, one might think that a sum
like (\ref{soh}) would be dominated by causets which in no way resemble
a spacetime, leading to a sort of ``entropy catastrophe''.
Nevertheless, it is not hard to forestall this catastrophe, and in fact
the most naive choice of stochastic dynamics already does so.  (Maybe this
is not so different from the situation in ordinary quantum
mechanics, where the smooth paths, which form a set of measure zero in
the space of all paths, are the ones which dominate the sum over histories
in the classical limit.)

The dynamics in question, which we will call ``transitive percolation'',
is perhaps the most obvious model of a randomly growing causet.  It is
an especially simple instance of a sequential growth dynamics, in which
each new element forges a causal bond independently with each existing
element with probability $p$, where $p\in[0,1]$ is a fixed parameter of
the model.  (Any causal relation implied by transitivity must then be
added in as well.)

{}From a more static perspective, one can also describe transitive
percolation by the following algorithm for generating a random poset:
\begin{enumerate}
\item Start with $n$ elements labeled $0,1,2,\cdots,n-1$ \  ($n=\infty$
 is not excluded.)
\item With a fixed probability $p$, introduce a relation between every
  pair of points labeled $i$ 
  and $j$, where $i<j$.
\item Form the transitive closure of these relations (e.g. if
  $2\prec5$ and $5\prec8$ then enforce that $2\prec8$.)
\end{enumerate}
Expressed in this manner, the model appears as a species of one
dimensional directed percolation; hence the name we have given it
(following D.~Meyer).

{}From a physical point of view, transitive percolation has some appealing
features, both as a model for a relatively small region of spacetime
and as a cosmological model for spacetime as a whole.  For
$p\sim{1/n}$, there is a percolation transition, where the causet goes
qualitatively from a large number of small disconnected universes for
$p<p_{\mathrm{crit}}$ to a single connected universe for
$p>p_{\mathrm{crit}}$.  Moreover, computer simulations suggest
strongly that the model possesses a continuum limit and exhibits
scaling behavior in that limit with $p$ scaling roughly like
$c\log{n}/n$ \cite{cont_lim,scaling}.  The ``cosmology'' of transitive
percolation is also suggestive --- the universe cycles endlessly
through phases of expansion, stasis, and contraction (via fluctuation)
back down to a single element \cite{posts}.

\begin{figure}[htbp]
\center
\scalebox{.6}{\includegraphics{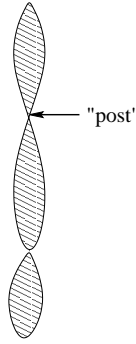}}
\caption{Transitive percolation cosmology}
\label{trans_perc_cos}
\end{figure}

{}From all this, it is clear that the causets generated by transitive
percolation do not at all resemble the 3-tier, generic causets of
Kleitman and Rothschild, but rather they have the potential to reproduce
a spacetime or a part of one.  
Nevertheless, the dynamics of transitive percolation is
not viable as a theory of quantum gravity.  One obvious reason is that
it is stochastic only in the purely classical sense, lacking quantum
interference.  Another reason is that the future of any element of the
causet is completely independent of anything ``spacelike related'' to
that element.  Therefore, the only spacetimes which a causal set
generated by transitive percolation could hope to resemble would be
homogeneous, such as the Minkowski or de Sitter spacetimes; but neither
of these possibilities is compatible with the periodic re-collapses
alluded to earlier.  At best, therefore, one could hope to reproduce a
small portion of such a homogeneous spacetime. 

On the other hand, in computer simulations of transitive percolation
\cite{AChR}, two independent (and coarse-graining invariant) dimension
estimators have tended to agree with each other, one such estimator
being that of \cite{meyer} and the other being a simple ``midpoint
scaling dimension''.  
(Some other indicators of manifold-like behavior
have tended to do much more poorly, but those are not invariant under
coarse graining, whereas one would in any case expect to observe
manifold like behavior only for a sufficiently coarse grained causal
set.)  In the pure percolation model, however, these dimension
indicators vary with time (i.e. with $n$) and one must rescale $p$ if
one wishes to hold the spacetime dimension constant.  One may ask,
then, if the model can be generalized by having $p$ vary with $n$ in
an appropriate sense.  We will see in the next section that something
rather like this is in fact possible.

The transitive percolation model, incidentally, has attracted the
interest of both mathematicians and physicists for reasons having
nothing to do with quantum gravity.  By physicists, it has been studied
as a problem in the statistical mechanical field of percolation, as we
have already alluded to.  By mathematicians, it has been studied
extensively as a branch of random graph theory (a poset being the same
thing as a transitive acyclic directed graph).
Some references on transitive percolation (viewed from whatever angle)
are 
\cite{brightwell,posts,bb,schulman,AChR,cont_lim,scaling}.

\section{Physical requirements on the dynamics}
\label{requirements}

As discussed in the previous section, one can think of transitive
percolation as a sort of ``birth process'', but as such, it is only
one special case drawn from a much larger universe of possibilities.
As preparation for describing these more general possible dynamical
rules, let us consider the growth-sequence of a causal set universe.

First element `0' appears (say with probability one, since the universe
exists).  Then element `1' appears, either related to `0' or not. Then
element `2' appears, either related to `0' or `1', or both, or neither.
Of course if $1\succ0$ and $2\succ1$ then $2\succ0$ by transitivity.
Then element `3' appears with some consistent set of ancestors, 
and so on and so forth.
Because of transitivity, each new element ends up with a partial stem of
the previous causet as its precursor set.
The result of this process, obviously, is a naturally labeled causet (finite
if we stop at some finite stage, or infinite if we do not) whose labels
record the order of succession of the individual births.
For illustration, consider the path in figure \ref{poscau}
delineated by the heavy arrows.  Along this path, 
element `0' appears initially, then 
element `1' appears to the future of element `0', then
element `2' appears to the future of element `0', 
but not to the future of `1', then 
element `3' appears unrelated to any existing element, then
element `4' appears to the future of 
elements `0', `1' (say, or `2', it doesn't matter) and `3', then
element `5' appears (not shown in the diagram), etc.

Let us emphasize once more that the labels 0, 1, 2, etc. are not
supposed to be physically significant.  Rather, the ``external time''
that they record is just a way to conceptualize the process, and any two
birth sequences related to each other by a permutation of their labels
are to be regarded as physically identical.

So far, we have been describing the kinematics of sequential growth.
In order to define a dynamics for it, 
we may give, 
for each $n$-element causet $C$, 
the {\it transition probability}
from it to each of its possible children.  
Equivalently, we give a transition probability
for each partial stem within $C$.  We wish to construct a general
theory for these transition probabilities by subjecting them to
certain natural conditions.  In other words, we want to construct the
most general (classically stochastic) ``sequential growth dynamics''
for causal sets.\footnote
{By choosing to specify our stochastic process in terms of transition
 probabilities, we have assumed in effect that the process is Markovian.
 Although this might seem to entail a loss of generality, the loss is
 only apparent, because the condition of discrete general covariance
 introduced below would have forced the Markov assumption on us,
 even if we had not already adopted it.}
In stating the following conditions, we will employ
the terminology introduced in the Introduction.

\subsubsection{The condition of internal temporality}

By this imposing sounding phrase, we mean simply that each element is
born either to the future of, or unrelated to, all existing elements;
that is, no element can arise to the past of an existing element.

We have already assumed this tacitly in describing what we mean by a
sequential growth dynamics.  An equivalent formulation is that the
labeling induced by the order of birth must be {\it natural}, as
defined above.  The logic behind the requirement of internal
temporality is that all physical time is that of the intrinsic order
defining the causal set itself.  For an element to be born to the past
of another would be contradictory: it would mean that an event
occurred ``before'' another which intrinsically preceded it.

\subsubsection{The condition of discrete general covariance}

As we have been emphasizing, the ``external time'' in which the causal
set grows (equivalently the induced labeling of the resulting poset)
is not meant to carry any physical information.  We interpret this in
the present context as being the condition that the net probability of
forming any particular $n$-element causet $C$ is independent of the
order of birth we attribute to its elements.  Thus, if
$\gamma$ is any path through the poset $\poscau$ of finite causal sets
that originates at the empty causet and terminates at $C$, then the
product of the transition probabilities along the links of $\gamma$
must be the same as for any other path arriving at $C$.  (So general
covariance in this setting is a type of path independence).
We should recall here, however, that, as observed earlier, a link in
$\poscau$ can sometimes represent more than one possible transition.
Thus our statement of path-independence, to be technically correct,
should say that the answer is the same no matter which transition
(partial stem) we select to represent the link.  
Obviously, this immediately entails that all
such representatives share the same transition probability.

We might with justice have required here conditions that are
apparently much stronger, including the condition that {\it any} two
paths through $\poscau$ with the same initial and final endpoints have
the same product of transition probabilities.  However, it is easy to
see that this already follows from the condition stated.\footnote
{If $\gamma$ does not start with the empty causet $C_0$, but at $C_s$,
we can extend it to start at $C_0$ by choosing any fixed path from $C_0$
to $C_s$.  Then different paths from $C_s$ to $C_e$ correspond to
different paths between $C_0$ and $C_e$, and the equality of net
probabilities for the latter implies the same thing for the former.}
We therefore do not make it part of our definition of discrete general
covariance, although we will be using it crucially.

\label{physical}
Finally, it is well to remark here that just because the ``arrival
probability at $C$'' is independent of path/labeling, that does not
necessarily mean that it carries an invariant meaning.  On the
contrary a statement like ``when the causet had 8 elements it was a
chain'' is itself meaningless before a certain birth order is chosen.
This, also, is an aspect of the gauge problem, but not one that
functions as a constraint on the transition probabilities that define
our dynamics.  Rather it limits the physically meaningful {\it
questions} that we can ask of the dynamics.  Technically, we expect
that our dynamics (like any stochastic process) can be interpreted as
a probability measure on a certain $\sigma$-algebra, and the
requirement of general covariance will then serve to select the
subalgebra of sets whose measures have direct physical meaning.

\subsubsection{The Bell causality condition}

The condition of ``internal temporality'' may be viewed as a very weak type
of causality condition.  The further causality condition we introduce now
is quite strong, being similar to that from which one derives Bell's
inequalities.  We believe that such a condition is appropriate for a
classical theory, and we expect that some analog will be valid in the
quantum case as well.  (On the other hand, we would have to abandon Bell
causality if our aim were to reproduce quantum effects from a classical
stochastic dynamics, as is sometimes advocated in the context of ``hidden
variable theories''.  Given the inherent non-locality of causal sets, 
there is no logical reason why such an attempt would have to fail.)

The physical idea behind our condition is that events occurring in some
part of a causal set $C$ should be influenced only by the portion of $C$
lying to their past.  In this way, the order relation constituting $C$
will be causal in the dynamical sense, and not only in name.  In terms of
our sequential growth dynamics, we make this precise as the requirement
that the ratio of the transition probabilities leading to two possible
children of a given causet depend only on the triad consisting of the
two corresponding precursor sets and their union.

Thus, let $C\rightarrow{}C_1$ designate a transition from $C\in\C_n$ to
$C_1\in\C_{n+1}$, and similarly for $C\rightarrow{}C_2$.  Then, the Bell
causality condition can be expressed as the equality of two 
ratios\footnote%
{In writing (\ref{BCr}), we assume for simplicity that both numerators
 and both denominators are nonzero, this being the only case we will
 have occasion to treat in the present paper.}:
\bne
  \frac {prob(C \rightarrow C_1)} {prob(C \rightarrow C_2)} 
   =
  \frac {prob(B \rightarrow B_1)} {prob(B \rightarrow B_2)} 
\label{BCr}
\ene
where 
$B\in\C_m$, ${m}\le{n}$, is the union of 
the precursor set of $C\rightarrow C_1$ with 
the precursor set of $C\rightarrow C_2$, 
$B_1\in\C_{m+1}$ is $B$ with an element added in the same manner as in 
the transition $C\rightarrow{C_1}$, and 
$B_2\in\C_{m+1}$ is $B$ with an element added in the same manner as in 
the transition $C\rightarrow C_2$.\footnote%
{Recall that the precursor set of the transition $C\to{}C_1$ is the
 subposet of $C$ that lies to the past of the new element that forms
 $C_1$.}
(Notice that if the union of the precursor sets is the
entire parent causet, then the Bell causality condition reduces to a
trivial identity.)

To clarify the relationships among the causets involved, it may help to
characterize 
the latter in yet another way.
Let $e_1$ be the element born in the transition $C\to{}C_1$ and
let $e_2$ be the element born in the transition $C\to{}C_2$.
Then
$C_i=C\union\{e_i\}$ ($i=1,2$), 
and we have
$B=(\past\,e_1)\union(\past\,e_2)$
and 
$B_i=B\union\{e_i\}$ ($i=1,2$).

\begin{figure}[htbp]
\center
\scalebox{.6}{\includegraphics{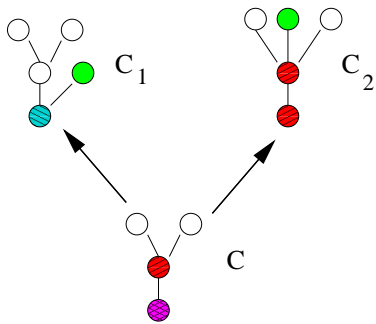}}
\caption{Illustrating Bell causality}
\label{bc_ex}
\end{figure}

\begin{figure}[htbp]
\center
\scalebox{.6}{\includegraphics{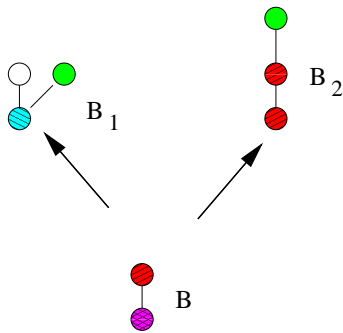}}
\caption{Illustrating Bell causality - spectators do not affect relative
probability}
\label{bc_ex2}
\end{figure}

By its definition, Bell causality relates ratios of transition
probabilities belonging to one ``stage'' of the growth process to ratios
of transition probabilities belonging to previous stages.  For
illustration,
consider the case depicted in figure \ref{bc_ex}.
The precursor $P_1$ of the transition $C\rightarrow{}C_1$ contains
only the earliest (minimum) element of $C$, shown in the figure as a
pattern-filled dot. 
The precursor $P_2$ of
$C\rightarrow{}C_2$ contains as well the next earliest element, shown
as a (different pattern)-filled dot.  
The union of the two
precursors is thus $B={P_1}\union{P_2}=P_2$.  The elements of $C$
depicted as
open dots belong to neither precursor.  Such elements will be called
\emph{spectators}.  Bell causality says that the spectators can be
deleted without affecting relative probabilities.  Thus the ratio of
the transition probabilities of figure \ref{bc_ex} is equal to that of
figure \ref{bc_ex2}.

\subsubsection{The Markov sum rule}

As with any Markov process, we must require that the sum of the full
set of transition probabilities issuing from a given causet be unity.
However, the set we have to sum over depends in a subtle manner on the
extent to which we regard causal set elements as ``distinguishable''.
Heretofore we have identified distinct transitions with distinct
precursor sets of the parent.  In doing so, we have in effect been
treating causet elements as distinguishable (by not identifying with
each other,
precursor sets related by automorphisms of the parent),
and this is what we shall continue to do.  Indeed, this is the
counting of children used implicitly by transitive percolation, so we
keep it here for consistency.  With respect to the diagram of figure
\ref{poscau}, this method of counting has the effect of introducing
coefficients into the sum rule, equal to the number of partial stems
of the parent which could be the precursor set of the transition.  For
the transitions depicted there, these coefficients (when not one) are
shown next to the corresponding arrow.\footnote%
{One might describe the result of setting these coefficients to unity as
 the case of ``indistinguishable causet elements''.  It appears that 
 in this case
 a dynamics with a richer structure obtains: instead of the transition
 probability depending only on the size of the precursor set and the
 number of its maximal elements, it is sensitive to more details of the
 precursor set's structure.}

We remark here that these sum-rule coefficients admit an alternative
description in terms of embeddings of the parent into the child (as a
partial stem).  Instead of saying ``the number of partial stems of the
parent which could be the past of the new element'', we could say ``the
number of order preserving injective maps from the parent onto partial
stems of the child, divided by the number of automorphisms of the
child''.  (The proof of this equivalence will appear in \cite{rideout}.)

\section{The general form of the transition probabilities}

We seek to derive a general prescription which gives, consistent with
our requirements, the transition probability from an element of
$\C_n$ to an element of $\C_{n+1}$.  To avoid having to deal with
special cases, we will assume throughout that no transition
probability vanishes.  Thus the solution we find may be termed
``generic'', but not absolutely general.

In this connection, we want to point out that one probably does not
obtain every 
possible solution of our conditions by taking limits of the generic
solution, and the special theories which result from taking certain
transition probabilities to vanish must be treated separately.\footnote%
{Indeed, the requirement of Bell causality itself must be given
 an unambiguous interpretation when some of the transition probabilities
 involved are zero.}
One such special theory is the \emph{originary percolation} model,
which is the same as the transitive percolation model, but with the
added restriction that each element except the original one must have
at least one ancestor among the previous elements.  The net effect is
that the growing causal set is required to have an ``origin'' (=
unique minimum element) at all stages.  (Generalizations are also
possible in which a more complex full stem of the causet is enforced.)
The poset of originary causets can be transformed into the poset of
all causets (exactly) by removing the origin from every originary
causet.  The transition probabilities for originary percolation are
just those of ordinary transitive percolation with an added factor of
$(1-q^n)$ in the denominator at stage $n$.

\subsection{Counting the free parameters}

A theory of the sort we are seeking provides a probability for each
transition, so without further restriction, it would contain a free
parameter for every possible antichain of every possible (finite)
causet.  We will see, however, that the requirements described above in
Section \ref{requirements} drastically limit this freedom.

\begin{lemma}
There is at most one free parameter per family.
\label{1param_family}
\end{lemma}

\noindent 
\textbf{Proof:} 
Consider a parent and its children.
Every such child, except
the timid child, participates in a Bell causality equation with the
gregarious child.  
(See the proof of Lemma \ref{bc_consis} in the Appendix.)  
Hence (since Bell causality equates ratios), all 
these transitions are determined up to an overall factor.
This leaves two free parameters for the family.  The Markov sum
rule gives another equation, which exhausts itself in determining
the probability of the timid child.  Hence precisely one
free parameter per family remains after Bell Causality
and the sum rule are imposed.  $\Box$

\begin{lemma}
The probability to add a completely disconnected element (the
``gregarious child transition'') depends only on the cardinality of the
parent causal set.
\label{q's}
\end{lemma}

\noindent \textbf{Proof:}
\begin{figure}[htbp]
\center
\scalebox{.9}{\includegraphics{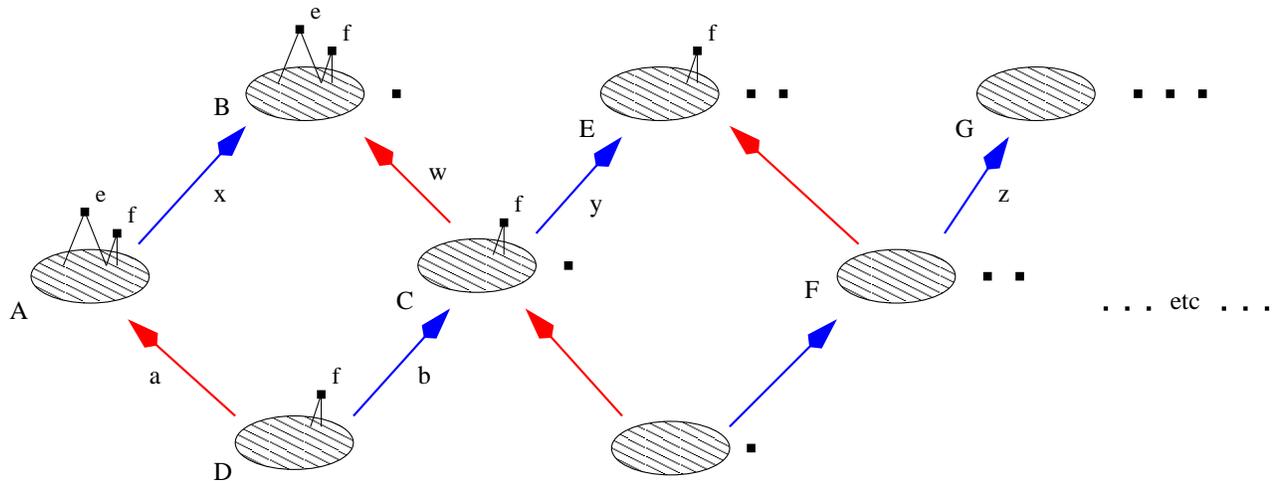}}
\caption{Equality of ``gregarious child'' transitions}
\label{GC}
\end{figure}
Consider an arbitrary causet $A$, with a maximal element $e$, as
indicated in figure \ref{GC}.  Adjoining a disconnected
element to $A$ produces the causet $B$.  Then, removing $e$ from $B$
leads to the causet $C$, which can be looked upon as the gregarious
child of the causet $D=A\backslash\{e\}$.  Adding another disconnected
element to $C$ leads to a causet $E$ with (at least) two completely
disconnected elements.  Now, by general covariance,
\be
      ax = bw 
\ee 
and by Bell causality,
\be 
    \frac{y}{w} = \frac{b}{a}
\ee 
(the disconnected element in $C$ acts as the spectator here).  Thus
\be 
    ax = bw = ay \Longrightarrow x=y
\ee 
(Recall that we have assumed that no transition probability vanishes.)
Repeating our deductions with $C$ in the place of $A$ in the above
argument (and a new maximal element $f$ in the place of $e$), we see
that $y=z$, where $z$ is the probability for the transition from $F$
to $G$ as shown.  Continuing in this way until we reach the antichain
$A_n$ shows finally that $x=q_n$, where we define $q_n$ as the
transition probability from the $n$-antichain to the
($n+1$)-antichain.  Since our starting causet $A$ was not chosen
specially, this completes the proof.  $\qed$

If our causal sets are regarded as entire universes, then a
gregarious child transition corresponds to the spawning of a new,
completely disconnected universe (which is not to say that this new
universe will not connect up with the existing universe in the
future).  Lemma \ref{q's} proves that the probability for this to
occur does not depend on the internal structure of the existing
universe, but only on its size, which seems eminently reasonable.
In the sequel, we will call this probability $q_n$.

With Lemmas 1 and 2, we have reduced the number of free parameters
(since every family has a gregarious child) to 1 per stage, or what is
the same thing, to one per causal set element.  In the next sections we
will see that no further reduction is possible based on our stated
conditions.  Thus, the transition probabilities $q_n$ can be identified
as the free parameters or ``coupling constants'' of the theory.  They
are, however, restricted further by inequalities that we will derive
below.

\subsection{The general transition probability in closed form}

Given the $q_n$, the remaining transition probabilities (for the
non-gregarious children) are determined by Bell causality and the sum
rule, as we have seen.  Here we
derive an expression in closed form for an arbitrary transition
probability in terms of causet invariants and the parameters $q_n$.

\subsubsection{Mathematical form of transition probabilities}

We use the following notation:

\smallskip
\begin{tabular}{|c|c|}
\hline
$\alpha_n$ & an arbitrary transition probability from $\C_n$ to $\C_{n+1}$\\
$\beta_n$  & a transition whose precursor set is not the entire parent
             (`non-timid' transition)\\
$\gamma_n$ & a transition whose precursor set \emph{is} the entire parent
             (`timid' transition)\\
\hline
\end{tabular}\\

\smallskip
\noindent
Notice that the subscript $n$ here refers only to the number of elements
of the parent causet; it does not exhibit which particular transition
of stage $n$ is intended.  A more complete notation might provide $\alpha$,
$\beta$ and $\gamma$ with further indices to specify both the parent
causet and the precursor set within the parent.

We also set $q_0\ideq{}1$ by convention.

\begin{lemma}
Each transition probability $\alpha_n$ of stage $n$ has the form
\bne
           q_n \sum_{i=0}^n {\lambda_i \frac{1}{q_i}}
\label{FN}
\ene
where the $\lambda_i$ are integers depending on the individual transition
in question.
\label{integer_form}
\end{lemma}

\noindent \textbf{Proof:} 
This is easily seen to be true for stages 0 and 1.  
Assume it is true for stage $n-1$.  
Consider a non-timid transition probability $\beta_n$ of stage $n$.  Bell
causality gives 
\be
     \frac{\beta_n}{q_n} = \frac{\alpha_{n-1}}{q_{n-1}}
\ee
where $\alpha_{n-1}$ is an appropriate transition probability from stage $n-1$.
So by induction
\bne 
  \beta_n = \alpha_{n-1} \frac{q_n}{q_{n-1}} 
  = \sum_{i=0}^{n-1} \lambda_i \frac{q_{n-1}}{q_i} \frac{q_n}{q_{n-1}} 
  = \sum_{i=0}^{n-1} \lambda_i \frac{q_n}{q_i} .
\label{BT}
\ene 
For a timid transition probability $\gamma_n$, we use the Markov sum rule: 
\bne
 \label{TT}
 \gamma_n = 1 - \sum_j  \beta_{nj}
\ene
where $j$ labels the possible non-timid transitions (i.e. the set of proper
partial stems of the parent).\footnote%
{Of course, more than one stem will in general correspond to the same
link in $\poscau$.  If we redefined $j$ to run over links in $\poscau$, then
(\ref{TT}) would read
$\gamma_n=1-\sum_j\chi_j\beta_{nj}$, where $\chi_j$ is the
``multiplicity'' of the $j$th link.}
But then, substituting (\ref{BT}) yields immediately
$$
  \gamma_n 
   = 1 - \sum\limits_j \sum\limits_{i=0}^{n-1} {\lambda_{ji}\over q_i} q_n
   = 1 - \sum\limits_{i=0}^{n-1} {\sum_j \lambda_{ji} \over q_i} q_n \,,
$$
which we clearly can put into the form (\ref{FN}) by taking
$\lambda_i=-\sum_j\lambda_{ji}$ for $i<n$ and
$\lambda_n=1$.                                                  $\Box$

\subsubsection{Another look at transitive percolation}
The transitive percolation model we described earlier is consistent
with the four conditions of Section 3.  To see this, consider an
arbitrary causal set $C_n$ of size $n$.  The transition probability
$\alpha_n$ from $C_n$ to a specified causet $C_{n+1}$ of size $n+1$ is
given by \bne
\label{pp}
\alpha_n = p^m (1-p)^{n-\pp} 
\ene 
where $m$ is the number of maximal elements in the precursor set  and
$\pp$ is the size of the entire precursor set.
(This becomes clear if one recalls how the precursor set of a newborn
element is generated in transitive percolation: first a set of ancestors
is selected at random, and then the ancestors implied by transitivity
are added.  From this, it follows immediately that a given stem
$S\subseteq{}C_n$ results from the procedure iff (i) every maximal
element of $S$ is selected in the first step, and (ii) no element of
$C_n\less{}S$  is selected  in the first step.)  
In particular, we see that the
``gregarious transition'' will occur with probability $q_n=q^n$, where
$q=1-p$. 

Now consider our four conditions.  Internal temporality was built in
from the outset, as we know.  Discrete general covariance is seen to
hold upon writing the net probability of a given $C_n$ explicitly in
terms of causet invariants (writing it in ``manifestly covariant
form'') as
$$
   P(C_n) = W  p^L  q^{{n\choose 2}-R}
$$
where $L$ is the number of links in $C_n$, $R$ the number of relations,
and $W$ the number of (natural) labelings of $C_n$. 

To see that transitive percolation obeys Bell causality, consider an
arbitrary parent causet.  The transition probability to a given child
is 
exhibited
in eq. (\ref{pp}).  Consider two different children, one with
$(m,\pp)$=($m_1$,$\pp_1$) and the other with
$(m,\pp)$=($m_2$,$\pp_2$).  
Bell causality requires that the ratio of their
transition probabilities be the same as if the parent were 
reduced to the
union of the precursor sets of the two transitions, i.e.
it requires
\be
\frac{p^{m_1} q^{n-\pp_1}}{p^{m_2} q^{n-\pp_2}} = 
	\frac{p^{m_1} q^{n'-\pp_1}}{p^{m_2} q^{n'-\pp_2}}
\ee
where $n'$ is the cardinality of the union of the precursor sets of the two
transitions.  
Thus, Bell causality is satisfied by inspection.

Finally, the Markov sum rule is essentially trivial.  At each stage of
the growth process, a preliminary choice of ancestors is made by a
well-defined probabilistic procedure, and each such choice is mapped
uniquely onto a choice of partial stem.  Thus the induced probabilities
of the partial stems sum automatically to unity.

\subsubsection{The general transition probability}

In the previous section we have shown that transitive percolation
produces transition probabilities 
(\ref{pp})
consistent with all our conditions.
By equating the right hand side of (\ref{pp}) to the general form
(\ref{FN}) of Lemma \ref{integer_form}, we can solve for the
$\lambda_i$ and thus obtain the general solution of our conditions:
\be
\alpha_n = \sum_{i=0}^n{\lambda_i \frac{1}{q_i}} \, q_n
	= p^m (1-p)^{n-\pp} 
	= (1-q)^m q^{n-\pp}
\ee
Expanding the factor $(1-q)^m$, and using the fact that $q_n=q^n$ for
transitive percolation, we get
\be
\lambda_i = (-)^{\pp-i} {m \choose \pp-i} .
\ee
So an arbitrary transition probability in the general dynamics is,
according to (\ref{FN})
\be
\alpha_n = \sum_{i=0}^n (-)^{\pp-i} {m \choose \pp-i} \frac{q_n}{q_i} .
\ee
Noting that the binomial coefficients are zero for $\pp-i \notin
\{0..m\}$, and rearranging the indices, 
we obtain
\bne
\label{gen_trans_prob}
\fbox{
$\displaystyle \alpha_n = 
	\sum_{k=0}^m (-)^k {m \choose k} \frac{q_n}{q_{\pp-k}}$ .}
\ene
This form for the transition probability exhibits its causal nature
particularly clearly: except for the overall normalization factor $q_n$,
$\alpha_n$ depends only on invariants of the associated precursor set.

\subsection{Inequalities}

Since the $\alpha_n$ are classical probabilities, each must lie between
0 and 1, and this in turn restricts the possible values of the $q_n$.
Here we show that it suffices to impose only one inequality per stage;
all the others (two per child) then follow.  More precisely, what we
show is that, if $q_n>0$ for all $n$, and if $\alpha_n\ge{0}$ for the
``timid'' transition from the $n$-antichain, then all the 
$\alpha_n$ lie in $[0,1]$.  This we establish in the following two
``Claims''.
\\[3mm]
\noindent \textbf{Claim} \,
\emph{In order that all the transition probabilities $\alpha_n$ fall between 0
and 1, it suffices that each timid transition probability be $\ge 0$.}
\\[3mm]
\noindent \textbf{Proof:}  
As described in the proofs of lemmas 1 and 5,
each non-timid transition (of stage $n$) is given (via Bell causality) by
\be
\alpha_n = \alpha_m \frac{q_n}{q_m}
\ee
where $m$ is some natural number less than $n$.  The $q$'s are
positive.  So if the probabilities of the previous stages are
positive, then the non-timid probabilities of stage $n$ are also positive.
It follows by induction that all but the timid transition probabilities
are positive (since $\alpha_0=q_0=1$ obviously is).  But for the timid
transition of each family, we have
\bne
  \gamma_n = 1 - \sum_i \beta_i
  \label{GF}
\ene
where each $\beta_i$ is positive.  If any of the $\beta_i$ is greater
than one, $\gamma_n$ will obviously be negative.  Also (\ref{GF})
plainly cannot be greater than one.  Consequently, if we
require that $\gamma_n$ be positive, then all transition probabilities
in the family will be in $[0,1]$.  $\Box$

In a timid transition, the entire parent is the precursor set, so $\pp=n$.
The inequalities constraining each probability 
of a given family
to be in $[0,1]$ therefore
reduce to the sole condition 
\bne
   \sum_{k=0}^m (-)^k {m \choose k} \frac{1}{q_{n-k}} \geq 0 \,.
\label{tnm}
\ene

\vspace{2mm}
\noindent \textbf{Claim} \,
\emph{The most restrictive inequality of stage $n$ is the one arising from the
$n$-antichain, i.e. the one for which $m=n$.  All other inequalities 
of stage $n$ follow from this inequality and the inequalities for
smaller $n$.}
\\[2mm]

\noindent \textbf{Proof:}
Assume that we have, for $m=n$, 
\be
\sum_{k=0}^n (-)^k {n \choose k} \frac{1}{q_{n-k}} \geq 0 .
\ee
Add to this the inequality from stage $n-1$,
\be
\sum_{k=0}^{n-1} (-)^k {n-1 \choose k} \frac{1}{q_{n-k-1}} =
\sum_{k=0}^{n} (-)^{k-1} {n-1 \choose k-1} \frac{1}{q_{n-k}} \geq 0
\ee
to get
\be
\sum_{k=0}^{n-1} (-)^k {n-1 \choose k} \frac{1}{q_{n-k}} \geq 0 .
\ee
This is the inequality of stage $n$ for $m=n-1$.  
(We have used
the identity ${n \choose k} = {n-1 \choose k} + {n-1 \choose k-1}$.)
Adding 
to it
the inequality of stage $n-1$ with $m=n-2$ yields the
inequality of stage $n$  for $m=n-2$.  
Repeating this process will give all the
inequalities of stage $n$. $\qed$

It is helpful to introduce the quantities
\bne
\fbox{ $ \displaystyle
  t_n = \sum_{k=0}^n (-)^{n-k} {n \choose k} \frac{1}{q_k}  $} 
\label{TD}
\ene
Obviously, we have $t_0=1$ (since $q_0=1$), and we have seen that the
full set of inequalities restricting the $q_n$ will be satisfied iff
$t_n\ge0$ for all $n$.  (Recall we are assuming $q_n>0, \, \forall n$.)
Moreover, given the $t_n$, we can recover the $q_n$ by inverting
(\ref{TD}): 
\begin{lemma}
\bne
\label{q_of_t}
  \fbox{ $ \displaystyle
  \frac{1}{q_n} = \sum_{k=0}^n {n \choose k} t_k$}
\ene
\end{lemma}
\noindent \textbf{Proof:}
This follows immediately from the identity
$$
  \sum\limits_{k=0}^n {n \choose k} (-)^{n-k} {k \choose m} = \delta^n_m
  \qquad\qquad\qquad \Box
$$
Thus, the $t_n$ may be treated as free parameters (subject only to
$t_n\ge{}0$ and $t_0=1$), and the $q_n$ can then be derived from 
(\ref{q_of_t}).
If this is done, 
the remaining transition probabilities $\alpha_n$ can be
re-expressed more simply in terms of the $t_n$ by inserting
(\ref{q_of_t}) into (\ref{gen_trans_prob}) to get
\be
\frac{\alpha_n}{q_n}  
     =  \sum_{l} t_l \sum_{k} (-)^k {m \choose k} {\pp-k \choose l}
     =  \sum_{l} t_l {\pp-m \choose \pp-l}
\ee
whence
\bne
\fbox{ 
 $ \displaystyle
  \alpha_n = \frac
             {\sum_{l=m}^{\pp} {\pp-m \choose \pp-l} t_l}
	     {\sum_{j=0}^n {n \choose j} t_j}
  $} 
\label{alpha_of_t}
\ene
Here, we have used an identity for binomial coefficients
that can be found on page 63 of \cite{Feller}.

In this way, we arrive at the general solution of our inequalities.
(Actually, we go slightly beyond our ``genericity'' assumption that
$\alpha_n\not=0$ if we allow some of the $t_n$ to vanish; but no harm is
done thereby.)

Let us conclude this section by noting that (\ref{q_of_t}) implies
\bne
  q_0 \equiv 1 \geq q_1 \geq q_2 \geq q_3 \geq \cdots
 \label{qin}
\ene
If we think of the $q_n$ as the basic parameters or ``coupling
constants'' of our sequential growth dynamics, then it is as if the
universe had a free choice of one parameter at each stage of the
process.  We thus get an ``evolving dynamical law'', but the evolution
is not absolutely free, since the allowable values of $q_n$ at every
stage are limited by the choices already made.  On the other hand, if
we think of the $t_n$ as the basic parameters, then the free choice is
unencumbered at each stage.  However, unlike the $q_n$, the $t_n$
cannot be identified with any dynamical transition probability.
Rather, they can be realized as ratios of two such probabilities,
namely as the ratio $x_n/q_n$, where $x_n$ is the transition
probability from an antichain of $n$ elements to the timid child of
that antichain.  (Thus, if we suppose that the evolving causet at the
beginning of stage $n$ is an antichain, then $t_n$ is the probability
that the next element will be born to the future of {\it every}
element, divided by the probability that the next element will be born
to the future of {\it no} element.)

\subsection{Proof that this dynamics obeys the physical requirements}
\label{obeys}

To complete our derivation, we must show that the sequential growth
dynamics given by (\ref{gen_trans_prob}) or (\ref{alpha_of_t}) obeys the
four conditions set out in section \ref{requirements}.

\subsubsection{Internal temporality}
This condition is built into our definition of the growth process.

\subsubsection{Discrete general covariance}
We have to show that the product of the transition probabilities
$\alpha_n$ associated with a labeling of a fixed finite causet $C$ is
independent of the labeling.  But this follows immediately from 
(\ref{gen_trans_prob}) [or (\ref{alpha_of_t})]
once we notice that what remains after the overall product
$$
  \prod\limits_{j=0}^{|C|-1}q_j
$$
is factored out, is a product over all elements $x\in{C}$ of poset
invariants depending only on the structure of $\past(x)$.

\subsubsection{Bell causality}
Bell causality states that the ratio of the transition probabilities
for two siblings depends only on the union of their precursors.
Looking at (\ref{gen_trans_prob}), consider the ratio of two
such probabilities
$\alpha_{n1}$ and $\alpha_{n2}$.  The $q_n$ factors will cancel,
leading to an expression which depends only on $\pp_1$, $\pp_2$, $m_1$,
and $m_2$.  Since these are all determined by the structure of the
precursor sets, Bell causality is satisfied.

\subsubsection{Markov sum rule}
The sum rule states that the sum of all transition probabilities
$\alpha_n$ from a given parent $C$ (of cardinality $|C|=n$) is unity.  
Since a child can
be identified with a partial stem of the parent,
we can write this condition, in view of (\ref{alpha_of_t}), as
\bne
\sum_S \sum_l t_l {|S|-m(S) \choose l-m(S)} = \sum_j t_j {n \choose j}
\label{Msr}
\ene
where $S$ ranges over the partial stems of $C$.
This must hold for any $t_l$, since they may be chosen
freely.  Reordering the sums and equating like terms yields
\bne
   \forall l, \ \sum_S {|S|-m(S) \choose l-m(S)} = {n \choose l} \,,
\label{identities}
\ene
an infinite set of identities which must hold if the sum rule
is to be satisfied by our dynamics.

The simplest way to see that (\ref{identities}) is true is to resort to
transitive percolation,
for which $t_l=t^l$, where $t=p/q=p/(1-p)$.
In that case we know that the sum rule is satisfied, so by inspection of
(\ref{Msr}), we see that the identity (\ref{identities}) must be true.

A more intuitive proof is illustrated well by
the case of $l=3$.  Group the terms on the left side 
according to the number of maximal elements:
\def\clt#1#2{#1\,|\,#2}
\be
\begin{array}{ccccccccr}
\sum\limits_{\clt{S}{m(S)=0}} {|S|-0 \choose 3-0} & + &
\sum\limits_{\clt{S}{m(S)=1}} {|S|-1 \choose 3-1} & + & 
\sum\limits_{\clt{S}{m(S)=2}} {|S|-2 \choose 3-2} & + & 
\sum\limits_{\clt{S}{m(S)=3}} {|S|-3 \choose 3-3} & = & {n \choose 3}\\
0 & + & 
\sum\limits_{\clt{S}{m(S)=1}} {|S|-1 \choose 2} & + &
\sum\limits_{\clt{S}{m(S)=2}} (|S|-2) & + &
\sum\limits_{\clt{S}{m(S)=3}} 1 & = & {n \choose 3}
\end{array}
\ee
The first term is zero because the only partial stem with zero maximal
elements is empty (i.e. $|S|=0$).  
The second term is a sum over all partial stems with one maximal
element.  This is equivalent to a sum over elements, with the element's
inclusive past forming the partial stem.  The summand chooses every
possible pair of elements to the past of the maximal element.  Thus the
second term overall counts the 3-element subcausets of $C$ with a
single maximal element.  There are two possibilities
here, the three-chain \chain3 and the ``lambda'' \wedge.  
The third term sums over partial stems with two maximal elements, 
which is equivalent to summing over 2 element antichains, 
the inclusive past of the antichain being the partial stem.  
The summand then counts the
number of elements to the past of the two maximal ones.  Thus the third
term overall 
counts the number of three element subcausets with
precisely two maximal elements.  Again there are two possibilities, the
``V'' \V, and the ``L'', \Lcauset.
Finally, the fourth term is a sum over partial stems with three maximal
elements, and this can be interpreted as a sum over all three element
antichains \3antichain.
As this example illustrates, then,
the left hand side of (\ref{identities}) counts the number of $l$ element
subcausets of $C$, placing them into ``bins'' according to the number
of maximal elements of the subcauset.  
Adding together the bin sizes yields the total number of $l$ element
subsets of $C$, which of course equals ${n \choose l}$. 

\subsection{Sample cosmologies}
\label{cosmologies}

The physical consequences of differing choices of the $t_n$ remain to
be explored.  To get an initial feel for this question, we list
 some simple examples.  
(Recall our convention that $t_0=1$, or equivalently, $q_0=1$, where
$q_0$ is the probability that the universe is born at all.\footnote%
{So, is the answer to the old question why something exists rather than
 nothing, simply that it is notationally more convenient for it to be so?})
\begin{itemize}
\item{``Dust universe''}
\be
     t_0=1, \ t_i = 0, \; i \geq 1
\ee
This universe is simply an antichain, since, according to (\ref{q_of_t}), 
$q_n=1$ for all $n$.

\item{``Forest universe''}
\be
       t_0 = t_1 = 1; \; t_i = 0, \; i \geq 2
\ee
This yields a universe consisting wholly of trees,
since (see the next example) $t_2=t_3=t_4=\cdots=0$ implies that no
element of the causet can have more than one past link.  
The particular choice of $t_1=1$ has in addition the remarkable property
that, as follows easily from (\ref{alpha_of_t}), every allowed transition
of stage $n$ has the same probability $1/(n+1)$.

\item{Case of limited number of past links}
\be
        t_i=0, \; i > n_0
\ee
Referring to expression 
(\ref{alpha_of_t})
one sees at once that $\alpha_n$ vanishes if $m>n_0$.  Hence, no element
can be born with more than $n_0$ past links or ``parents''.  This means
in particular that any realistic choice of parameters will have $t_n>0$
for all $n$, since an element of a causal set
faithfully embeddable in Minkowski space
would have an infinite number of past links. 

\item{Transitive percolation}
\be
             t_n = t^n
\ee
We have seen that for
transitive percolation, 
$q_n=q^n$, where $q=1-p$.  Using the
binomial theorem, it is easy to learn from 
(\ref{q_of_t}) 
or (\ref{TD}) that this choice of $q_n$ corresponds to $t_n=t^n$ with
$t=p/q$.  Clearly, $t$ runs from 0 to $\infty$ as $p$ runs from 0 to 1.

\item{A more lifelike choice?}
\bne
        t_n = \frac{1}{n!}
\label{lifelike}
\ene
We have seen that transitive percolation with constant $p$ yields
causets which could reproduce --- at best --- only limited portions of
Minkowski space.  To do any better, one would have to scale $p$ so that
it decreased with increasing $n$
\cite{cont_lim,scaling,AChR}.
This suggests that $t_n$ should fall off faster than in any percolation
model, hence (by the last example) faster than exponentially in $n$.
Obviously, there are many possibilities of this sort
(e.g. $t_n\sim{}e^{-\alpha{}n^2}$), but one of the simplest is
$t_n\sim{c}/n!$  \ This would be our candidate of the moment for a
physically most realistic choice of parameters.

\end{itemize}

\section{The stochastic growth process as such}
\label{class_measure}
We have seen that, associated with every {\it labeled} causet
$\tilde{C}$ of size $N$, is a net ``probability of formation''
$P(\tilde{C})$ which is the product of the transition probabilities
$\alpha_i$ of the individual births described by the labeling:
$$
   P(\tilde{C}) = \prod\limits_{i=0}^{N-1} \alpha_i
$$
where $\alpha_i=\alpha(i,\pp_i,m_i)$ is given by 
(\ref{gen_trans_prob}) or (\ref{alpha_of_t}).  
We have also
seen that $P$ is in fact independent of the labeling and may be written
as $P(C)$ where $C$ is the unlabeled causet corresponding to
$\tilde{C}$.  To bring this out more clearly, let us define
\bne
 \label{lambda-d}
 \lambda(\pp,m) = \sum\limits_{k=m}^\pp {\pp-m\choose \pp-k} t_k
\ene
Then $q_i=\lambda(i,0)^{-1}$ and we have
$$
  \alpha_i(i,\pp_i,m_i) = {\lambda(\pp_i,m_i)\over\lambda(i,0)} \,,
$$
whence
$ P(\tilde{C}) = \prod\limits_{i=0}^{N-1} \lambda(\pp_i,m_i) / 
                 \prod\limits_{j=0}^{N-1} \lambda(j,0)$,
or
expressed more intrinsically,
\bne
\label{PC0}
        P(C) = {
                \prod\limits_{x\in C} \lambda(\pp(x),m(x)) 
                \over
                \prod\limits_{j=0}^{|C|-1} \lambda(j,0) } \,,
\ene
where $\varpi(x)=|\past{x}|$ and $m(x)=|{\rm maximal}(\past x)|$.
This expression, as far as it goes, is manifestly ``causal'' and
``covariant'' in the senses explained above.  As also explained above,
however, it has no direct physical meaning.
Here we briefly discuss
some probabilities which {\it do} have 
a fully
covariant meaning and show how,
in simple cases, they are related to $N\to\infty$ limits of
probabilities like (\ref{PC0}).

\def\Prob{{\rm Prob}}

First, let us notice that the net probability of arriving at a
particular $C\in\poscau$ is not $P(C)$ but
$$
  \Prob_N(C) = W(C) \, P(C)
$$
where $N=|C|$ and $W(C)$ is the number of inequivalent\footnote%
{Two labelings of $C$ are equivalent iff related by an automorphism of $C$.}
labelings of $C$, or in other words, the total number of paths through
$\poscau$ that arrive at $C$, each link being taken with its proper
multiplicity. 

Now as a rudimentary example of a truly covariant question, let us take 
``Does the two-chain ever occur as a partial stem of $C$?''.
The answer to this question will be a probability, $P$, which 
it is natural to identify as
$$
         P = \lim_{N\to\infty} \Prob_N (X_N)  \,,
$$
where $X_N$ is the event that ``at stage $N$'', $C$ possesses a partial
stem which is a two-chain.    
In this connection, 
we conjecture that the
questions of the form
``Does $P$ occur as a partial stem of $C$?'' furnish a
physically complete set, when $P$ ranges over all (isomorphism
equivalence classes of) finite causets.

\section{Two Ising-like state-models}
\label{Ising}

In this section, we present two Ising-like state-models from which
$P(C)$ of equation (\ref{PC0}) can be obtained.  In the main we just
indicate the results, leaving the details to appear elsewhere
\cite{rideout}.
The two models come from taking (\ref{gen_trans_prob}) or,
respectively, (\ref{alpha_of_t}) as the starting point.  In each case,
the idea is to interpret the binomial coefficients which occur in
these formulas as describing a sum over subsets of relations of $C$.
If we work with (\ref{gen_trans_prob}) these will be subsets of the
set of {\it links} of $C$; if we work with (\ref{alpha_of_t}) they
will be subsets of the set of relations of $C$ that are {\it not}
links.

Let us take first equation  (\ref{gen_trans_prob}).  
Reinterpreting the binomial coefficients in the manner indicated, and
proceeding as in the derivation of (\ref{PC0}), we arrive at an
expression for $P(C)$ in terms of a sum over
$\Integers_2$-valued ``spins'' $\sigma$ living on the relations of
$C$.  In summing over configurations, however, the spins $\sigma$ on
the non-link relations are set permanently to 1; only those on the
links vary.  With $\sigma=1$ interpreted as ``presence'' and
$\sigma=0$ as ``absence'', the contribution of a particular spin
configuration $\sigma$ is an overall sign times the product of one
``vertex factor'' for each $x\in{C}$.  The vertex factor is
$q_r^{-1}$, where $r$ is the number of present relations having $x$ as
future endpoint, and the sign is $(-)^a$, where $a$ is the number of
absent relations.  (In addition, there is a constant overall factor 
in $P(C)$ of $\prod_{j=0}^{|C|-1}q_j$.)

In the second state model, 
we begin with (\ref{alpha_of_t}), 
or better (\ref{PC0}) itself, 
and proceed similarly.  
The result is again a sum over spins $\sigma$ 
residing on the relations, 
this time 
with all the terms being positive
(as is required of physical Boltzmann weights).
%(in closer agreement with ordinary spin systems).  
In this second model, 
the spins on the links are set permanently to 1
while those on the non-links vary.   
The ``vertex factor'' coming from $x\in{C}$ 
now is $t_r$, 
where $r$ is again 
the number of relations 
present and ``pointing to $x$''.

These two models (and especially the second) show that our sequential
growth dynamics can be viewed as a form of ``induced gravity''
obtained by summing over (``integrating out'') the values of our
underlying spin variables $\sigma$.  This underlying ``matter'' theory
may or may not be physically reasonable (Does it obey its own version
of Bell causality, for example?  Is it local in an appropriate sense?),
but at a minimum, it serves to 
illustrate how a theory of non-gravitational matter can be hidden
within a theory that one might think to be limited to gravity
alone.\footnote%
{In this connection, it bears remembering that Ising matter can
 produce fermionic as well as bosonic fields, at least in certain
 circumstances. \cite{id89,ple97}}
\footnote%
{References \cite{kaz} and \cite{staud} 
(for which we thank an anonymous referee)
describe
a similar example of ``hidden'' matter fields
in the context of 2-dimensional random surfaces 
(Euclidean signature quantum gravity) 
and the associated matrix models 
in the continuum limit.  
Unfortunately, the matter fields used 
(Ising spins or ``hard dimers'') 
were unphysical in the sense that the partition function 
was a sum of Boltzmann weights 
which were not in general real and positive.  
This is much like our first state model described above.  
To the extent that the analogy between 
these two, rather different, situations 
holds good, 
our results here suggest that there might be, 
in addition to the matter fields employed in \cite{staud},
another set of fields
with physical choices of the coupling constants,
which could reproduce the same effective dynamics for
the random surface.}

\section{Further Work}

Our dynamics can be simulated; 
for $t_n=1/n!$ it takes a minute or so to generate a 64 element causet
on a DEC Alpha 600 workstation.
Analytic results, so far, are available only for the special case of
transitive percolation.  
An important question, of course, is whether some choice
of the $t_n$ can reproduce general relativity, or at least reproduce a
Lorentzian manifold for some range of $t$'s and of $n=|C|$.
Similarly, one can ask whether our ``Ising matter'' gives rise to an
interesting effective field theory and what relation it has with the
local scalar matter on a background causal set studied in
\cite{daughton,salgado}?

Another set of questions concerns the possibility of a more ``manifestly
covariant'' formulation of our sequential growth dynamics -- or of more
general forms of causal set dynamics.  Can Bell causality be formulated
in a gauge invariant manner, without reference to a choice of birth
sequence?  Is our conjecture correct that all meaningful assertions are
logical combinations of assertions about the occurrence of partial
stems (``past sets'')?  
(Such questions seem likely to arise with special
urgency in any attempt to generalize our dynamics to the quantum case.)

Also, there are the special cases we left unstudied.  There exist
originary analogs of all of our dynamics, for example.  Are there other
special, non-generic cases of interest?

We might continue multiplying questions, but let's finish with the
question of how to discover a quantum generalization of our dynamics.
Since our theory is formulated as a type of Markov process, and since a
Markov process mathematically is a probability measure on a suitable
sample space, the natural quantum generalization would seem to be a
{\it quantum} measure\cite{qmeasure} (or equivalent ``decoherence
functional'') on the same sample space.
The question then would be whether
one could find appropriate quantum analogs of Bell causality and general
covariance formulated in terms of such a quantum measure.
If so, we could hope that, just as in the classical case treated herein,
these two principles would lead us to a relatively unique quantum
causet dynamics,\footnote
{See \cite{criscuolo} for a promising first step toward such a dynamics.}
or rather to a family of them among which a potential
quantum theory of gravity would be recognizable.

It is a pleasure to thank Avner Ash for a stimulating discussion 
at a critical stage of our work.
The research reported here was supported in part by NSF grant
PHY-9600620 and by a grant from the Office of Research and Computing of
Syracuse University.

\vspace{9mm}
\noindent
\textbf{\Large Appendix: Consistency of the conditions}
\vspace{5mm}
\label{consistency}

\noindent
Our analysis of the conditions of Bell causality et al. unfolded
in the form of several lemmas.  Here we present some similar lemmas
which strictly speaking are not needed in the present context, but which
further elucidate the relationships among our conditions.  We expect
these lemmas can be useful in any attempt to formulate generalizations
of our scheme, in particular quantal generalizations.

\begin{lemma}
The Bell causality equations are mutually consistent.
\label{bc_consis}
\end{lemma}

\begin{figure}[htbp]
\center
\scalebox{.9}{\includegraphics{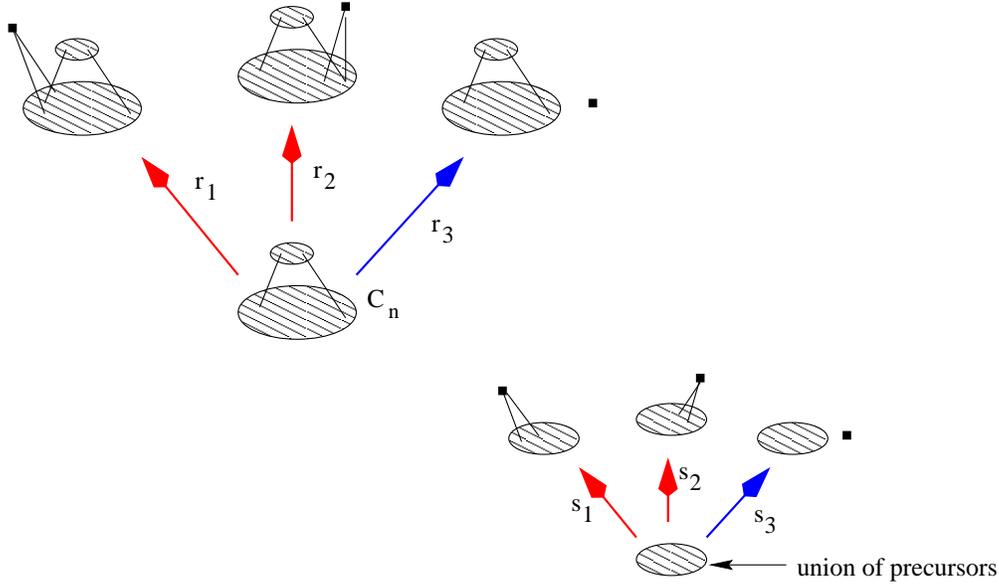}}
\caption{two families related by Bell causality}
\label{BCconsis}
\end{figure}

\noindent \textbf{Proof:} 
The top of figure \ref{BCconsis} shows three children of an arbitrary
causal set $C_n$.  The shaded ellipses represent  portions of
$C_n$.  The small square indicates the new element whose birth
transforms $C_n$ into a causal set $C_{n+1}$ of the next stage.  The
smaller ellipse ``stacked on top of'' the larger ellipse represents a
subcauset of $C_n$ which does not intersect the precursor set of any
of the transitions being considered (i.e. none of its elements lie to
the past of any of the new elements).  This small ellipse thus consists
entirely of ``spectators'' to the transitions under consideration.  The
bottom part of figure \ref{BCconsis} shows the corresponding parent and
children when these spectators are removed.

Notice that one of the three children is the gregarious child.  We
will show that the Bell causality equations between
this child and each of the others imply all remaining Bell causality
equations within this family.  Since no Bell causality equation
reaches outside a single family (and since, within a family, the Bell
causality equations that involve the gregarious child obviously always
possess a solution --- in fact they determine all ratios of transition
probabilities except for that to the timid child), this will prove the
lemma.

In the figure $t_1$ and $t_2$ represent a general pair of
transitions related by a Bell causality equation, namely
\bne 
    \frac{t_1}{t_2} = \frac{s_1}{s_2} \,.
\label{BCE}
\ene 
But, as illustrated,
each of these is also related by a Bell causality equation to the
gregarious child, to wit:
\bne 
    \frac{t_1}{t_3} = \frac{s_1}{s_3} \quad \mbox{and} \quad 
    \frac{t_2}{t_3} = \frac{s_2}{s_3} 
\label{BCE2}
\ene
Since (\ref{BCE}) follows immediately from (\ref{BCE2}),
no inconsistencies can arise at stage $n$, and
the lemma follows by induction on $n$. $\Box$

\begin{lemma}
Given Bell causality and the further
consequences of general covariance that are
embodied in Lemma \ref{q's}, all the remaining general covariance
equations reduce to identities, i.e. they place no further restriction
on the parameters of the theory.
\end{lemma}

\noindent \textbf{Proof:}
\begin{figure}[htbp]
\center
\scalebox{.9}{\includegraphics{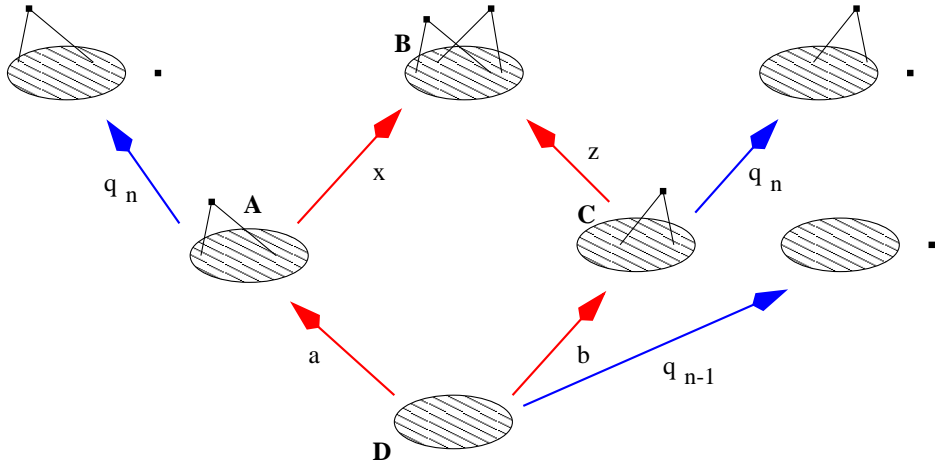}}
\caption{\label{consis} consistency of remaining general covariance conditions}
\end{figure}
Discrete general covariance states that the probability of forming a
causet is independent of the order in which the elements arise,
i.e. it is independent of the corresponding path through the poset of
finite causets.  

Now, general covariance relations always can be taken
to come from `diamonds' in the poset of causets, for the following
reason.  
As illustrated in figure \ref{consis}, any two parents $A$, $C$ of a
causet $B$ will have a common parent $D$ (a ``grandparent'' of $B$)
obtained by removing two suitable elements from $B$.  (By assumption $B$
must contain an element whose removal yields $A$ and another element
whose removal yields $C$.  Then remove these two elements.  For example,
consider the case where \wedgeo is the grandchild and it has the parent
\3antichain (by removing the maximal element of the wedge) and the parent
\wedge (by removing the disconnected element).  To find the
grandparent \twoach
%(a 2-antichain) 
remove both maximal elements from \wedgeo.)

% As illustrated in figure \ref{consis}, any pair of children
% of a causet (siblings) will have a common child obtained by adjoining
% both new elements of the two siblings, i.e. adding to the
% ``grandparent'' both the new element which defines one sibling and the
% new element which defines the other sibling.
% (For example, consider the case where the 2-antichain is the grandparent
% and it has the child \3antichain
% (by adding a disconnected element) 
% and the child \wedge
% (by adding an element to the future of both elements of the 2-antichain).  
% To find their common child \wedgeo add a disconnected element to \wedge,
% or an element to the future of two of the elements of \3antichain.)

Now, still referring to figure \ref{consis}, 
let $|D|=n$ and suppose inductively that all the general covariance
relations are satisfied up through stage $n$.  A new condition arising
at stage $n+1$ says that some path arriving at $B$ via $x$ has the
same probability as some other path arriving via $z$.  But, by our
inductive assumption, each of these paths can be modified to go
through $D$ without affecting its probability.  Thus, the equality of
our two path probabilities reduces simply to $ax=bz$.

Now by Bell causality and lemma \ref{q's},
\be
   \frac{x}{q_n} = \frac{b}{q_{n-1}} \,,
\ee
whence
\be
   ax = ab \frac{q_n}{q_{n-1}} \,.
\ee
But by symmetry, we also have
\be
   bz = ba \frac{q_n}{q_{n-1}}  \,;
\ee
therefore $ax=bz$, as required.   $\Box$

%: references 


\begin{thebibliography}{10}


\bibitem{qmeasure}
Rafael~D. Sorkin.
\newblock Quantum mechanics as quantum measure theory.
\newblock {\em Modern Physics Letters A}, 9(33):3119--3127, 1994.
\newblock $<$e-print archive: gr-qc/9401003$>$.

\bibitem{JFKK}
J.L. Friedman and A.~Higuchi.
\newblock State vectors in higher-dimensional gravity with quantum numbers of
  quarks and leptons.
\newblock {\em Nuclear Physics B}, 339:491--515, 1990.

\bibitem{causets0}
Rafael~D. Sorkin.
\newblock Forks in the road, on the way to quantum gravity.
\newblock {\em Int. J. Th. Phys.}, 36:2759--2781, 1997.
\newblock talk given at the conference entitled ``Directions in General
  Relativity'', held at College Park, Maryland, May, 1993. $<$e-print archive:
  gr-qc/9706002$>$.

\bibitem{causets1}
Rafael~D. Sorkin.
\newblock Spacetime and causal sets.
\newblock In J.~C. D'Olivo, E.~Nahmad-Achar, M.~Rosenbaum, M.P. Ryan, L.F.
  Urrutia, and F.~Zertuche, editors, {\em Relativity and Gravitation: Classical
  and Quantum}, pages 150--173, Singapore, December 1991. World Scientific.
\newblock (Proceedings of the {\it SILARG VII Conference}, held Cocoyoc,
  Mexico, December, 1990).

\bibitem{causets2}
Luca Bombelli, Joohan Lee, David Meyer, and Rafael~D. Sorkin.
\newblock Space-time as a causal set.
\newblock {\em Physical Review Letters}, 59:521--524, 1987.

\bibitem{causets4}
David Reid.
\newblock Introduction to causal sets: an alternative view of spacetime
  structure.
\newblock preprint, 1999.

\bibitem{bom87}
Luca Bombelli.
\newblock {\em Spacetime as a Causal Set}.
\newblock PhD thesis, Syracuse University, December 1987.

\bibitem{fot}
F.~Markopoulou and L.~Smolin,
\newblock ``Causal evolution of spin networks'',
\newblock {\it Nuc. Phys.} {\bf B508:} 409-430 (1997)
\newblock \eprint{gr-qc/9702025}

\bibitem{amb1}
J.~Ambj{\o}rn and R.~Loll,
\newblock  ``Non-perturbative Lorentzian quantum gravity, causality 
   and topology change'',
\newblock {\em Nuc. Phys. B} {\bf 536}: 407-434 (1999)
\newblock \eprint{hep-th/9805108}

\bibitem{amb2}
\newblock J.~Ambj{\o}rn, K.N.~Anagnostopoulos and R.~Loll,
\newblock ``A New Perspective on Matter Coupling in 2d Quantum Gravity'',
\newblock \eprint{hep-th/9904012}

\bibitem{brightwell}
Graham Brightwell.
\newblock Models of random partial orders.
\newblock In Keele, editor, {\em Surveys in combinatorics}, volume 187 of {\em
  London Math. Soc. Lecture Note Ser.}, pages 53--83. Cambridge University
  Press, Cambridge, 1993.

\bibitem{cont_lim}
David~P. Rideout and Rafael~D. Sorkin.
\newblock Continuum limit of percolated causal sets.
\newblock (in preparation).

\bibitem{scaling}
David~P. Rideout and Rafael~D. Sorkin.
\newblock Scaling behavior of percolated causal sets.
\newblock (in preparation).

\bibitem{posts}
B\'ela Bollob\'as and Graham Brightwell.
\newblock The structure of random graph orders.
\newblock {\em Siam J. Discrete Math}, 10(2):318--335, May 1997.

\bibitem{AChR}
Alan Daughton, Rafael~D. Sorkin, and C.R. Stephens.
\newblock Percolation and causal sets: A toy model of quantum gravity.
\newblock (in preparation).

\bibitem{meyer}
David~A. Meyer.
\newblock {\em The Dimension of Causal Sets}.
\newblock PhD thesis, Massachusetts Institute of Technology, 1988.

\bibitem{bb}
B\'ela Bollob\'as and Graham Brightwell.
\newblock Graphs whose every transitive orientation contains almost every
  relation.
\newblock {\em Israel Journal of Mathematics}, 59(1):112--128, 1987.

\bibitem{schulman}
C.~M. Newman and L.~S. Schulman.
\newblock One-dimensional $1/|j-i|^s$ percolation models: the existence of a
  transition for $s\leq2$.
\newblock {\em Comm. Math. Phys.}, 104(4):547--571, 1986.

\bibitem{rideout}
David~P. Rideout.
\newblock {\em Causal Set Dynamics}.
\newblock PhD thesis, Syracuse University, 1999.
\newblock (in preparation).

\bibitem{Feller}
William Feller.
\newblock {\em An Introduction to Probability Theory and Its Applications},
  volume~I.
\newblock Wiley, 1957.

\bibitem{id89}
Claude Itzykson and Jean-Michel Drouffe.
\newblock {\em Statistical Field Theory}, volume~2.
\newblock Cambridge University Press, 1989.

\bibitem{ple97}
V.~N. Plechko.
\newblock Anticommuting integrals and fermionic field theories for
  two-dimensional {I}sing models.
\newblock August 1997.
\newblock $<$e-print archive: hep-th/9607053$>$.

\bibitem{kaz}
V.A.~Kazakov,
\newblock ``The appearance of matter fields from quantum fluctuations of
    2D-gravity'', 
\newblock {\it Mod. Phys. Lett. A} {\bf 4:}: 2125-2139 (1989)

\bibitem{staud}
Matthias Staudacher,
\newblock ``The Yang-Lee edge singularity on a dynamical planar random
   surface'', 
\newblock {\it Nuc. Phys.} {\bf B336:} 349-362 (1990)


\bibitem{daughton}
Alan Daughton.
\newblock {\em The Recovery of Locality for Causal Sets and Related Topics}.
\newblock PhD thesis, Syracuse University, 1993.

\bibitem{salgado}
Roberto Salgado.
\newblock PhD thesis, Syracuse University, 1999.
\newblock (in preparation).

\bibitem{criscuolo}
A.~Criscuolo and H.~Waelbroeck.
\newblock Causal set dynamics: A toy model.
\newblock 1998.
\newblock $<$e-print archive: gr-qc/9811088$>$.

\end{thebibliography}
\end{document}